\documentclass[jmp,amsmath,amssymb,preprint]{revtex4-1}
\usepackage{dcolumn}
\usepackage{bm}
\usepackage{color}
\usepackage{rotating}
\usepackage{graphicx}
\usepackage{latexsym}
\usepackage{amssymb}
\usepackage{amsfonts}
\usepackage{soul}
\usepackage{color}
\usepackage{amsmath}

\begin{document}


\title{Thermoelectric spin voltage in graphene}

\author{Juan F. Sierra$^1$}
\email{juan.sierra@icn2.cat}
\author{Ingmar Neumann$^{1,2}$}
\author{Jo Cuppens$^{1}$}
\author{Bart Raes$^{1}$}
\author{Marius V. Costache$^1$}
\author{Sergio O. Valenzuela$^{1,3}$}
\email{SOV@icrea.cat}
\affiliation{$^1$Catalan Institute of Nanoscience and Nanotechnology (ICN2), CSIC and The Barcelona Institute of Science and Technology (BIST), Campus UAB, Bellaterra, 08193 Barcelona, Spain}
\affiliation{$^2$Universitat Auton\'oma de Barcelona, Bellaterra, 08193 Barcelona, Spain}
\affiliation{$^3$Instituci\'{o} Catalana de Recerca i Estudis Avan\c{c}ats (ICREA), 08070 Barcelona, Spain}


\date{\today}

\begin{abstract}

\textbf{
In recent years, new spin-dependent thermal effects have been discovered in ferromagnets, stimulating a growing interest in spin caloritronics, a field that exploits the interaction between spin and heat currents \cite{johnson1987,bauer2012}. Amongst the most intriguing phenomena is the spin Seebeck effect \cite{uchida2008,uchida2010,jaworski2010}, in which a thermal gradient gives rise to spin currents that are detected via the inverse spin Hall effect \cite{SOV2006,saitoh2006,SOV2015}. Non-magnetic materials such as graphene are also relevant for spin caloritronics, thanks to efficient spin transport \cite{tombros2007,han2014,SR2014}, energy-dependent carrier mobility and unique density of states \cite{veramarun2012,castroneto2009}. Here, we propose and demonstrate that a carrier thermal gradient in a graphene lateral spin valve can lead to a large increase of the spin voltage in the vicinity of the graphene charge neutrality point. Such an increase results from a thermoelectric spin voltage, which is analogous to the voltage in a thermocouple and that can be enhanced by the presence of hot carriers generated by an applied current \cite{berciaud2010,betz2012a,betz2012b,JFS2015}. These results can prove crucial to drive graphene spintronic devices and, in particular, to sustain pure spin signals with thermal gradients and to tune the remote spin accumulation by varying the spin-injection bias.}

\end{abstract}

\maketitle

To introduce the notion of the thermoelectric spin voltage (TSV), it is useful to adapt the analogy with conventional thermocouples from the original scenario that was proposed to explain the spin-Seebeck effect \cite{uchida2008}. In graphene, thermoelectric effects are sensitive to changes in the carrier density $n$. The Seebeck coefficient $S(n)$, schematically shown in Fig. 1a, changes sign across the charge neutrality point (CNP), when the dominant carriers change from electrons to holes or vice versa \cite{zuev2009,wei2009}. Therefore, a thermocouple can be fabricated using two sheets of this material with unequal $n$ (Fig. 1b). For carrier densities $n_1$ and $n_2$, the thermoelectric voltage is $V_S=-\Delta S\Delta T$, where $\Delta T$ is the temperature difference between the two ends of the sheets and $\Delta S = [S_2(n_{2}) - S_1(n_{1})]$.

In the spin-transport case, carriers can be considered as flowing in independent spin channels characterized by two spin sub-bands whenever the relevant length scale is smaller than the spin diffusion length $\lambda_{s}$. In graphene, $\lambda_{s}$ can exceed a few micrometres, while spin injection from a ferromagnet can result in a significant difference $\Delta n=n^{\uparrow}-n^{\downarrow}$ between the carrier densities for spin-down ($n^{\downarrow}$) and spin-up ($n^{\uparrow}$). This difference is reflected in the spin accumulation $\Delta \mu_0=\mu^{\uparrow}-\mu^{\downarrow}$, where $\mu^{\downarrow,\uparrow}$ are the spin-dependent electrochemical potentials associated to $n^{\downarrow,\uparrow}$ (Fig. 1c). Because $n^{\downarrow} \neq n^{\uparrow}$, the Seebeck coefficients for spin-down and spin-up carriers $S^{\downarrow,\uparrow} = S(n^{\downarrow,\uparrow})$ can also be different. By replacing $n_{1,2}$ with $n^{\downarrow, \uparrow}$ and $S_{1,2}$ with $S^{\downarrow,\uparrow}$, it becomes apparent that, within $\lambda_{s}$, the scenario sketched in Fig. 1b is intrinsically present in a single graphene sheet. The change in $\Delta \mu_0$, labelled $\delta\mu$, results in a thermoelectric spin voltage $\delta\mu/e \sim -(S^{\uparrow}-S^{\downarrow})\Delta T$ with $e$ the electron charge.
This analogy suggests that when $S^{\uparrow}<S^{\downarrow}$ the spin accumulation $\Delta \mu=\Delta \mu_0+\delta\mu$ at a remote detector can be enhanced by the presence of a temperature gradient.

In order to observe the TSV, we implement multi-terminal graphene devices comprising normal metal and spin sensitive electrodes as shown in Fig. 1d (see Methods, Supplementary Fig. 1). Spin injection and detection is achieved with ferromagnetic electrodes 3 and 4, respectively, which delimit the spin channel \cite{johnson1985,SOV2009}. Metal electrodes 1 and 2, and the graphene in between, define a heater that generates the temperature gradient in the spin channel and induces the TSV in electrode 4. A key aspect of the device geometry is that it allows independent control of the heat and spin sources and, therefore, separate thermal and spin injection effects.
As discussed below, it can also discriminate between the spin thermoelectric effect discussed here from the thermal spin injection, which is due to a thermal gradient in the injector and the spin-dependent Seebeck effect \cite{slachter2010,erekhinsky2012}.

We start by estimating the graphene Seebeck coefficient $S$ from the square resistance $R$ vs $n$ using the Mott relation \cite{zuev2009,wei2009},
$S_\mathrm{Mott}= \frac{\pi^2 k_{\mathrm{B}}^2T}{3e}\frac{d\mathrm{ln}R}{d\mu}\mid_{\mu=\mu_{\mathrm{F}}}$, where $k_{\mathrm{B}}$ is the Boltzmann constant and $\mu_{\mathrm{F}}$ the Fermi energy. The Mott Seebeck coefficient $S_\mathrm{Mott}$, shown in Figs. 2a (room temperature, RT) and 2b (77 K), is known to be a good approximation of $S$, both in magnitude and temperature dependence \cite{zuev2009,JFS2015}, and allows us to determine $\Delta T$ from the thermoelectric voltage.

To compare $S_\mathrm{Mott}$ with the actual thermoelectric response of the device, the voltage in the spin channel, $V_{dc}$, is measured vs the dc current, $I_{dc}$, in the heater. Figures 2c and 2d show the results at RT and at 77 K, respectively. In both cases, we observe a change from an upward ($V_{dc}>0$) to a downward ($V_{dc}<0)$ response with $n$, as expected from a thermoelectric effect.
Since the graphene is used as part of the heater, both lattice and carriers are simultaneously heated. However, the carrier temperature $T$ can differ from, and be much larger than, the lattice temperature \cite{berciaud2010,betz2012a,betz2012b}. Hot carriers propagate away from the heater region, but at RT they thermalize with the lattice before reaching the voltage electrodes \cite{JFS2015}. This leads to a parabolic dependence of $V_{dc} = -S \Delta T = -\Sigma I_{dc}^2$, owing to the classical relationship between temperature increase and Joule heating, $\Delta T = \alpha I_{dc}^2$. Here $\alpha$ is a constant that depends on the heater resistance, the geometry, and the thermal conductivity of the device components and substrate. The coefficient $\Sigma$ is extracted from Fig. 2c and is represented with symbols in Fig. 2a at different $n$. Consistent with the thermoelectric origin of $V_{dc}$, $\Sigma \approx \alpha S_\mathrm{Mott}$, with $\alpha \approx 430$ K/mA$^2$, which implies that $\Delta T \approx 1$ K at $I_{dc}=50$ $\mu$A.

At 77 K, the presence of hot carriers leads to a much larger temperature gradient than at RT and to the break-down of the classical relationship between Joule heating and $\Delta T$ \cite{JFS2015}. Despite that $S$ increases monotonically with $T$ \cite{zuev2009,wei2009}, $V_{dc}$ at 77 K is larger than at RT by an order of magnitude, while the parabolic dependence of $V_{dc}$ vs $I_{dc}$ transforms into a characteristic V-shape. The hot carriers obey a thermal distribution \cite{berciaud2010} but $\Delta T$ is difficult to model and depends strongly on $n$ \cite{betz2012b,JFS2015}. However, noting that $S \propto T$, we can estimate $\Delta T$ from the ratio between $S_\mathrm{Mott}$ and $V_{dc}$ in Fig. 2d. At the CNP a more precise estimation can be obtained from the slopes of $S_{Mott}$ and $V_{dc}$ vs $n$; we find that $\Delta T \sim 60$ K (CNP) and $\Delta T \sim 30$ K ($n= 10^{12}$ cm$^{-2}$), which is significantly larger than $\Delta T \approx 1$ K at RT. These large $\Delta T$ in the hot-carrier regime are highly favorable to observe the TSV.

Next, we investigate the graphene spin transport properties when an ac current $i_{ac}$ is applied in the non-local configuration, using electrodes 3 and 4 as spin injector and detector, respectively (Fig. 1d). Figure 2e shows the non-local resistance, $R_{NL}=V_{ac}/i_{ac}$, as a magnetic field $B$ along the long axis of the electrodes is swept to generate parallel ($\uparrow\uparrow$) and antiparallel ($\downarrow\uparrow$) magnetization alignments. The non-local spin signal, defined as $\Delta R_{NL} = R_{NL}^{\uparrow\uparrow}- R_{NL}^{\downarrow\uparrow}$, is about 1.5 $\Omega$. Figure 2f shows spin precession measurements with out-of-plane $B$, from which we determine the spin relaxation length $\lambda_{s}$, the spin relaxation time $\tau_{s}$, as well as the effective polarization $P$ of the electrodes (Supplementary Fig. 2).

Having characterized the spin transport and thermoelectric properties, the device is now biased in the configuration shown in Fig. 3a, where $I_{dc}$ and $i_{ac}$ are simultaneously applied. The current $I_{dc}$ generates a thermal gradient in graphene that induces the TSV, which is quantified from the change $\delta R^{S}_{NL}=\frac{P\delta\mu}{ei_{ac}}$ in the non-local signal $\Delta R_{NL}$. Note that $I_{dc}$ also generates a thermal gradient in the spin injector. Since $i_{ac}$ passes through the heater such gradient is modulated by it, therefore, an additional change $\delta R^{th}_{NL}$, associated with thermal spin injection, should be expected in $\Delta R_{NL}$. Nevertheless, as explained in the Supplementary Information, $\delta R^S_{NL}$ and $\delta R^{th}_{NL}$ are even and odd functions of $I_{dc}$, respectively, providing a straightforward procedure to disentangle them: $\delta R^S_{NL}$ is obtained from the average between the measurements for $\pm I_{dc}$, while $\delta R^{th}_{NL}$ is obtained by calculating half of the difference between the same measurements.

Figures 3b and 3c show $R_{NL}$ vs in-plane and out-of-plane $B$, respectively, for $I_{dc} = 0$ and 50 $\mu$A. In Fig. 3b, we observe a clear increase in $\Delta R_{NL}$ when the heater current is applied. An increase in $\Delta R_{NL}$ must originate from $\delta R^{S}_{NL}$ and/or $\delta R^{th}_{NL}$. Because $I_{dc}$ is injected in non-magnetic electrodes and the detection is non-local, the increase is of thermal origin, however, it cannot be explained by a simple increase in the graphene temperature since $\Delta R_{NL}$ weakly decreases with temperature (Supplementary Fig. 2). Similar conclusions are drawn from Fig. 3c, which also shows that the overall thermal effect is largest about the CNP.

Figure 3d shows $\delta R^{S}_{NL}$ vs $n$, as extracted from the measurements acquired with $|I_{dc}|=50$ $\mu$A; $\delta R^{th}_{NL}$ from the same measurements is shown in Supplementary Fig. 4. It is observed that $\delta R^{S}_{NL}$ displays a maximum value of $\approx 0.3$ $\Omega$ at the CNP, which represents a significant $\sim 20$\% increase in $\Delta R_{NL}$; $\delta R^{S}_{NL}$ also presents an electron-hole asymmetry, it decreases away from the CNP and clearly changes sign in the hole side for $|n|= 7 \times 10^{11} $ cm$^{-2}$. To understand these results, we first consider the scheme in Fig. 4. As depicted in Fig. 4a, the spin accumulation, leading to $n^{\downarrow} \neq n^{\uparrow}$, effectively shifts $S^{\downarrow}$ and $S^{\uparrow}$. As explained in the introduction, $\delta\mu \sim -e(S^{\uparrow}-S^{\downarrow})\Delta T$. Therefore, for small spin accumulation (typically $\Delta n$ is below 10$^{11}$ cm$^{-2}$), $\delta \mu$ can be approximated by $\delta\mu \sim -e\frac{\partial (S\Delta T)}{\partial n}\Delta n $, where we assume that, when hot carriers are involved, both $S$ and $\Delta T$ can depend on $n$. If $\Delta T$ was weakly dependent on $n$, then $\delta\mu \propto\frac{\partial S}{\partial n}$, which qualitatively describes $\delta R^{S}_{NL}\propto \delta \mu$ and suggests that the change in sign in $\delta R^{S}_{NL}$ roughly occurs when $S$ reaches an extreme (compare Figs. 3d and 4b).

A quantitative comparison with the experiment can be made from $\delta R^{S}_{NL}=\frac{P\delta\mu}{ei_{ac}}\sim -P\frac{\partial (S\Delta T)}{\partial n}\frac{\Delta n}{i_{ac}}$. Here, $\frac{\partial (S\Delta T)}{\partial n}$ is directly extracted from Fig. 2b, and $\Delta n$ estimated from the spin injection rate $\frac{P i_{ac}}{e}$, $\tau_s$, $\lambda_s$ and the width $w$ of the graphene as $\Delta n \sim \frac{P \tau_s}{2 e \lambda_s w} i_{ac}$. Using values at the CNP: $P=6$ \%, $\tau_s=250$ ps, and $\lambda_s=1$ $\mu$m, we find $\frac{\Delta n}{i_{ac}} \sim 3.6 \times 10^9$ cm$^{-2}$ $\mu$A$^{-1}$. Combining with $\frac{\partial (S\Delta T)}{\partial n} \sim 2.4\times 10^{-9}$ $\mu$V cm$^{-2}$, we obtain $\delta R^{S}_{NL}\sim 0.5$ $\Omega$, which is in reasonable agreement with the experimentally found value. Moreover, the modelled $\delta R^{S}_{NL}$ vs $n$, obtained by differentiating $S\Delta T$  (solid line in Fig. 3d), successfully reproduces all of the trends in the data, including the electron-hole asymmetry and the change in sign at $|n|= 7 \times 10^{11} $ cm$^{-2}$. Such an agreement conveys confidence to the interpretation of $\delta R^{S}_{NL}$ as arising from the TSV. The somewhat smaller experimental $\delta R^{S}_{NL}$ at the CNP could be a consequence of an overestimation of $\Delta n$ at the injector when the temperature gradient is applied.
Note that because the carrier cooling times are dependent on $n$ \cite{JFS2015,song2012}, the temperature for spin-up and spin-down carriers can be different, as observed in a nanopillar spin valve subjected to a heat current \cite{dejene2013}. This is considered in the model, as it estimates $\delta R^{S}_{NL}$ from the product between $S(n)$ and $\Delta T(n)$.

The spin splitting induced by the thermoelectric spin voltage in graphene is at least two orders of magnitude larger than that deriving from the spin-dependent Seebeck effect in metals \cite{slachter2010}. This striking result is a consequence of a relatively large Seebeck coefficient, its strong variation at the Fermi level, and the low density of states of graphene. It also stems from the  temperature gradient induced by hot carriers; the lattice temperature gradient is about an order of magnitude smaller and cannot explain the magnitude of the observed effect. Because the peak Seebeck coefficient scales as $ 1/\sqrt{n_r}$, the thermoelectric spin voltage can be enhanced by decreasing the residual carrier concentration $n_r$. The temperature gradient can also increase if the spin injector was part of the heater, resulting in a strong bias dependence of the device performance even at room temperature; however, other effects, such as a bias dependence of the polarization, could play a role in this case \cite{han2009,IN2014,gurram2017}. The demonstrated phenomenon is analogous to the enhancement, away from the CNP, generated with a drift current \cite{inglaaynes2016} but based on a thermal drift in combination with an $n$-dependent Seebeck coefficient. These observations are not exclusive to graphene and could be relevant for other materials presenting an energy-dependent conductance, for instance common semiconductors \cite{kikkawa1999}, or topological insulators, which exhibit exotic spin properties and strong thermoelectric effects. They can therefore lead to advances in spin caloritronics, where spin currents are controlled or sustained over long distances using heat currents.

\newpage

\section{Methods}

\textbf{Sample fabrication.} The devices are fabricated using monolayer graphene obtained by mechanical exfoliation from a highly oriented pyrolitic graphite source. The flakes are deposited on a p$^+$Si/SiO$_2$ substrate with a 440-nm SiO$_2$ layer. Monolayer flakes are identified by optical contrast analysis and the absorption of the optical light by the flakes, which was calibrated by Raman spectroscopy. The metallic and ferromagnetic (FM) electrodes were defined by two electron-beam lithography steps using a MMA/PMMA mask. In the first step, metallic electrodes are defined and 1 nm of Ti and 25 nm of Pd deposited using electron-beam evaporation in a chamber with a base pressure of 10$^{-8}$ Torr. In the second step, the FM electrodes are defined. TiO$_x$ barriers are fabricated in order to suppress the effect of the impedance mismatch between the ferromagnet and graphene. The fabrication of the barriers consisted in the evaporation of  4 $\textup{\r{A}}$ of Ti and then oxidation with pure oxygen during 30 min at a pressure $\sim$ 10$^{-2}$ Torr in the evaporation chamber. This evaporation/oxidation process was carried out twice in order to better control the quality of the TiO$_x$. The full process results in approximately 1 nm thick TiO$_x$ barriers, after which 30 nm of Co is deposited. The contact resistance is of the order of 10 k$\Omega$. The widths of the FM electrodes (100 nm and 120 nm) determine their coercive fields and allows us to control the relative configuration of their magnetizations (parallel and antiparallel). The distance between heater and spin injector (electrode 2 to electrode 3) is 1.2 $\mu$m. The length of the spin channel (electrode 3 to electrode 4) is 2.2 $\mu$m. A back-gate voltage applied to the p$^+$Si substrate is used to control the graphene carrier density $n$.

\textbf{Data availability}

The data that support the findings of this study are available from the corresponding authors upon reasonable request.

\vspace{5mm}

\noindent \textbf{Acknowledgments} We thank D. Torres for help in designing Fig. 1. This research was partially supported by the European Research Council under Grant Agreement No. 308023 SPINBOUND, by the European Union's Horizon 2020 research and innovation programme under grant agreement No. 696656 (Graphene Flagship), by the Spanish Ministry of Economy and Competitiveness, MINECO (under Contracts No. MAT2013-46785-P, No MAT2016-75952-R and Severo Ochoa No. SEV-2013-0295), and by
the CERCA Programme and the Secretariat for Universities and Research, Knowledge Department of the Generalitat de Catalunya 2014 SGR 56. J.F.S. and M.V.C. acknowledge support from MINECO Juan de la Cierva and Ram\'{o}n y Cajal program, respectively and J.C. from Generalitat de Catalunya, Beatriu de Pinos program.

\vspace{5mm}

\noindent \textbf{Author contributions} J.F.S., I.N. and S.O.V. planned the measurements. J.F.S. fabricated the samples and performed the experiments. J.C., B.R. and M.V.C. provided support for the device fabrication and M.V.C. for the measurements. J.F.S. and S.O.V. analyzed the data and wrote the manuscript. All authors discussed the results and commented on the manuscript. S.O.V supervised the work.
\vspace{5mm}

\noindent \textbf{Additional Information} Supplementary Information is available in the online version of the paper. Reprints and permissions information is available online at http://npg.nature.com/reprints. Correspondence and request for materials should be addressed to J.F.S. and S.O.V.
\newpage

\begin{figure}[ht]
\vspace{-10mm}\includegraphics[width=14cm]{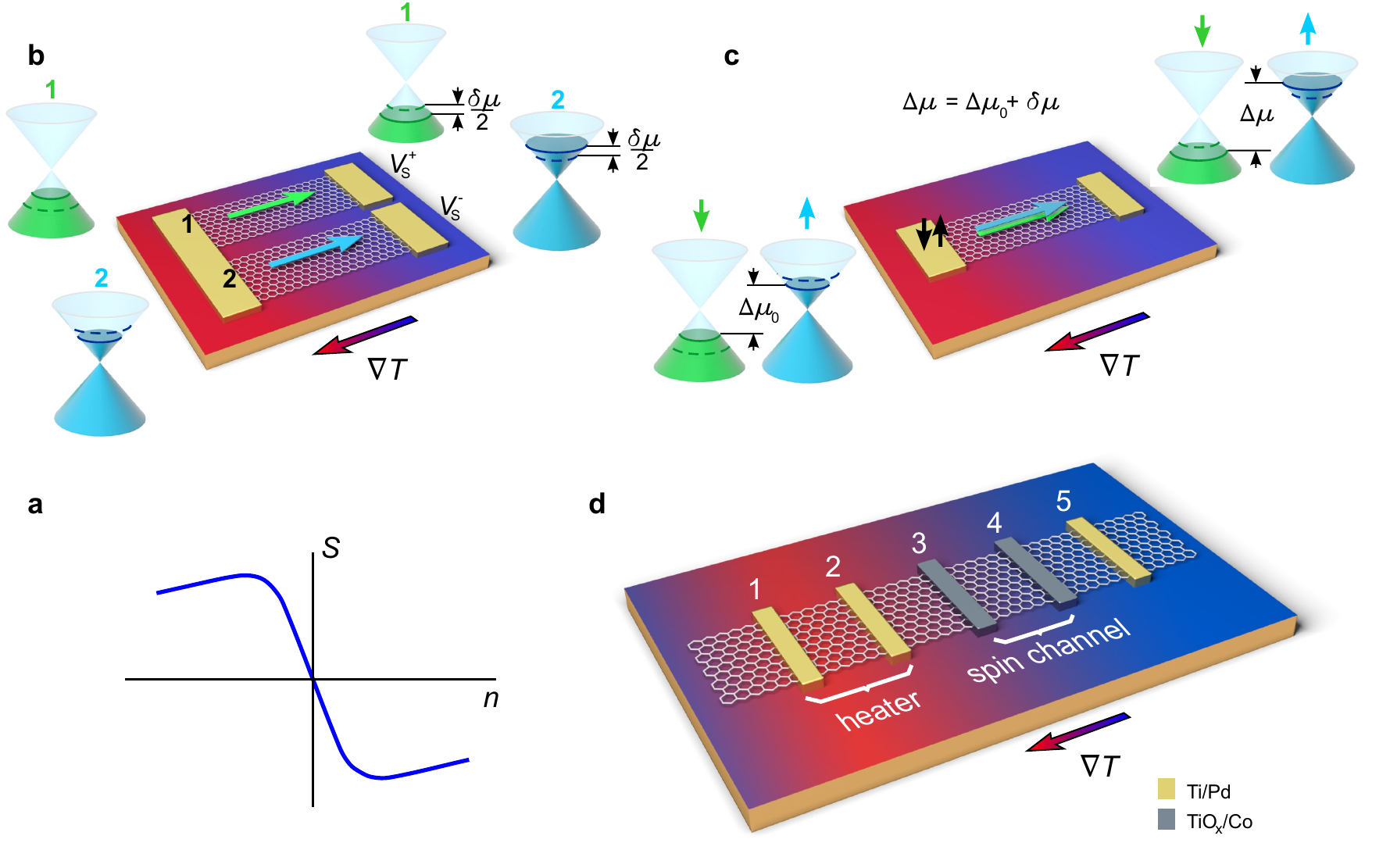}
\vspace{-2mm}
\caption{\textbf{Thermoelectric spin voltage (TSV)}. \textbf{a}, Qualitatively representation of the Seebeck coefficient $S$ in graphene about the charge neutrality point (CNP). \textbf{b}, Conventional thermocouple comprising two graphene sheets, 1 and 2, with carrier densities $n_1$ and $n_2$, and thus different Seebeck coefficient $S_1$ and $S_2$.  A thermoelectric voltage $V_{S} = V_S^+-V_S^-=-(S_2-S_1)\Delta T$ is built-up due to the temperature difference $\Delta T$ between the cold and the hot side of the sheets, which derives from the temperature gradient $\nabla T$. For the case drawn, $n_1=-n_2$ and thus $V_S=\delta\mu/e$. The flow direction of the majority carriers in 1 (holes) and 2 (electrons) is shown with green and blue arrows, respectively. \textbf{c}, At length scales smaller than the spin relaxation length, carriers with opposite spins belong to two independent transport channels. When the spin accumulation $\Delta \mu_0\neq0$, $S$ becomes spin dependent. A thermoelectric effect analogous to that in \textbf{b} leads to a TSV and a remote increase (decrease) of the spin accumulation $\Delta \mu=\Delta \mu_0+\delta \mu$ at the cold end. For simplicity, the case $n^{\downarrow}=-n^{\uparrow}$ is drawn. The flow direction of the carriers for spin-down and spin-up sub-bands is shown with green and blue arrows, respectively. \textbf{d}, Device configuration to detect the TSV proposed in \textbf{c}. A heating current between electrodes 1 and 2 generates $\nabla T$. Spin injection and detection is achieved with electrodes 3 and 4, which define the spin channel. The temperature difference in the spin channel generates the TSV that leads to a change in the spin signal measured in 4. The color of the substrate in \textbf{b-d} represents the temperature of the carriers in graphene.}
\label{Fig1}
\end{figure}

\begin{figure}[ht]
\includegraphics[width=11cm]{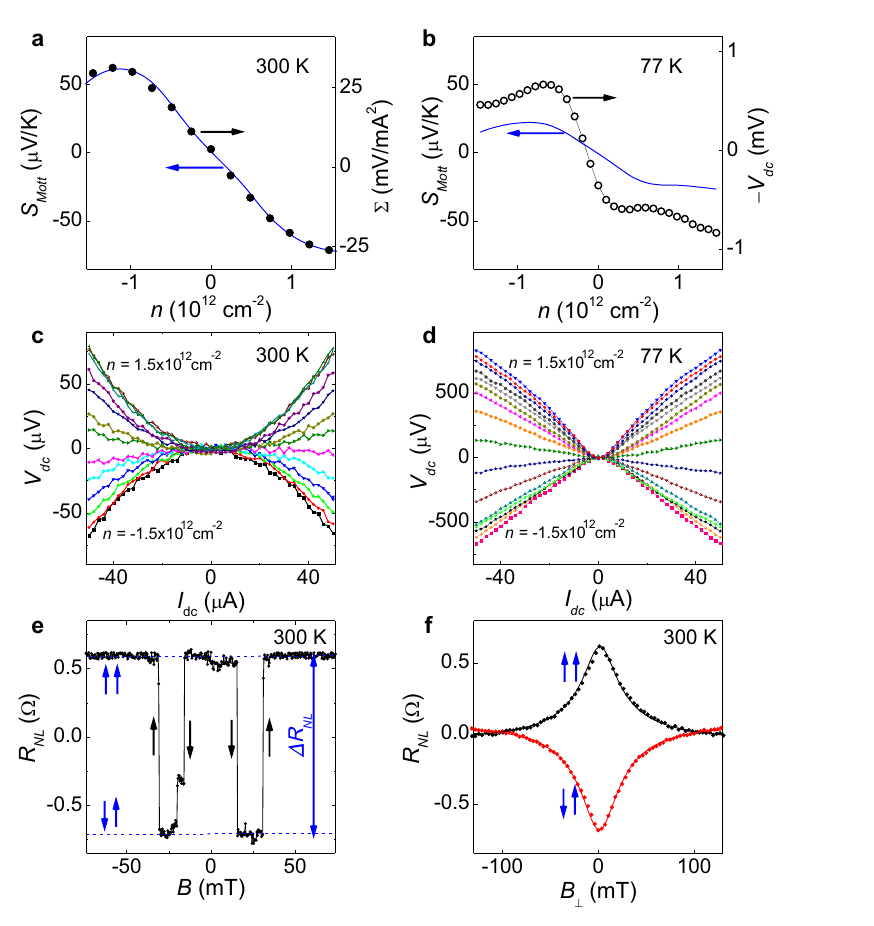}
\vspace{-3mm}
\caption{\textbf{Device characteristics}. \textbf{a}, Comparison between the Mott Seebeck coefficient $S_{Mott}$ vs carrier density $n$ (blue line and left axis) obtained from the graphene square resistance $R$ at room temperature (Supplementary Fig. 1) and the quadratic fitting coefficient $\Sigma$ vs $n$ (solid symbols and right axis) obtained from the thermoelectric measurements shown in \textbf{c}. \textbf{b}, $S_{Mott}$ vs $n$ (blue line and left axis) obtained from the graphene square resistance $R$ at 77 K (Supplementary Fig. 1). Thermoelectric voltage $V_{dc}$ vs $n$ in electrodes 3 and 4 (open symbols and right axis) for a heating current $I_{dc}=50$ $\mu$A applied between electrodes 1 and 2 (see Fig. 1d). \textbf{c, d}, $V_{dc}$ vs $I_{dc}$ at different $n$ at room temperature (\textbf{c}) and at 77 K (\textbf{d}). \textbf{e}, Typical non-local spin resistance $R_{NL}$ as a function of the magnetic field $B$ along the long axis of the ferromagnetic electrodes 3 and 4. The black arrows indicate the sweep direction. \textbf{f}, Spin precession measurements. Typical $R_{NL}$ as a function of an out-of-plane magnetic field $B_{\perp}$ for parallel (black) and antiparallel (red) configuration. The blue arrows in \textbf{e} and \textbf{f} indicate the relative orientation of the electrode magnetizations.}
\label{Fig2}
\end{figure}

\begin{figure}[ht]
\includegraphics[width=14cm]{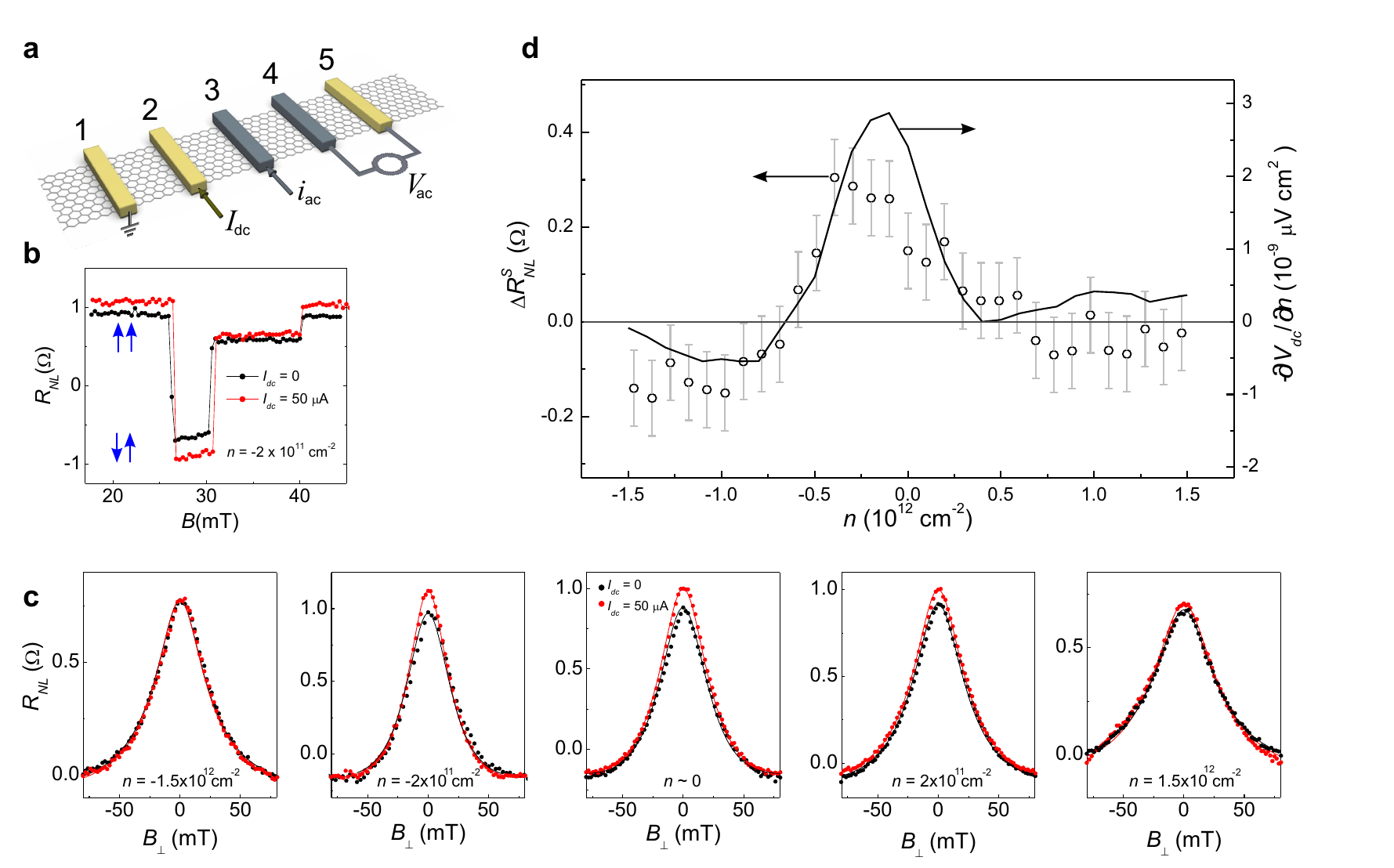}
\vspace{-3mm}
\caption{\textbf{Thermoelectric spin voltage detection}. \textbf{a}, Measurement configuration to detect the TSV. The dc heating current $I_{dc}$ and the ac spin injection current $i_{ac}$ are applied simultaneously, while the spin voltage $V_{ac}$ is measured at the frequency characterizing $i_{ac}$; $|i_{ac}| \ll |I_{dc}|$. \textbf{b}, Non-local spin resistance $R_{NL}$ vs magnetic field $B$ along the magnetization of the electrodes for $I_{dc}$ = 0 (black) and 50 $\mu$A (red). The step between 30 and 40 mT is due to a two-step switching of the ferromagnet. The blue arrows indicate the relative orientation of the ferromagnet magnetizations. \textbf{c}, $R_{NL}$ vs out-of-plane magnetic field $B_{\perp}$ at different $n$ for $I_{dc}$ = 0 (black) and 50 $\mu$A (red). The magnetization of the electrodes are in parallel configuration. In \textbf{b} and \textbf{c}, an increase of $R_{NL}$ is observed when $I_{dc}$ is applied; for clarity, a small spin-independent background has been subtracted for $I_{dc}$ = 50 $\mu$A. \textbf{d}, Change in $R_{NL}$ induced by the TSV, $\delta R^S_{NL}$, as a function of $n$ (open symbols and left axis).  The comparison to $\frac{\partial V_{dc}}{\partial n}$ at $I_{dc}$ = 50 $\mu$A (line and right axis) is suggested by theoretical modelling (Fig. 4). The data in \textbf{d} are obtained at fixed gate from the difference in spin signal for the parallel and antiparallel magnetization configuration of the electrodes at $B=0$; the error bars reflect the uncertainty in $R_{NL}$. Note that in \textbf{c} the positive contribution due to the thermal spin injection (Supplementary Fig. 4) suppresses the overall change induced by $I_{dc}$ at $n=\pm 1.5 \times 10^{12}$ cm$^{-2}$.}
\label{Fig3}
\end{figure}

\begin{figure}[ht]
\includegraphics[width=10cm]{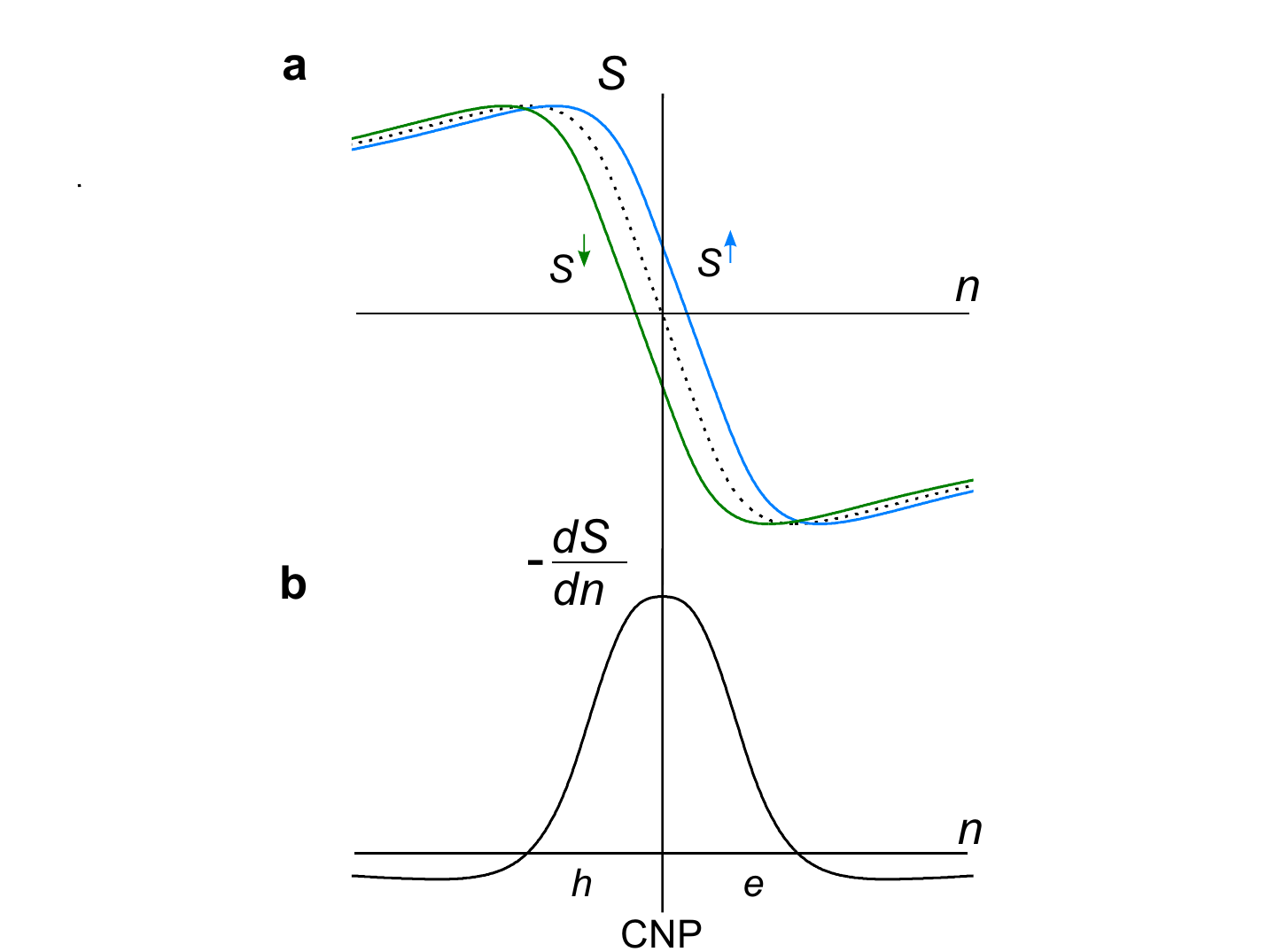}
\vspace{-3mm}
\caption{\textbf{Modelling and roles of $S$ and the spin accumulation}. Qualitatively representation of $S$ (\textbf{a}) and its derivative $dS/dn$ (\textbf{b}) about the CNP. $S$ is positive for holes ($h$) and negative for electrons ($e$). The spin accumulation, which is quantified by different $n$ for spin-down ($\downarrow$) and spin-up ($\uparrow$) sub-bands, results in a shift along $n$ of their respective Seebeck coefficients $S^{\downarrow}$ and $S^{\uparrow}$.  This leads to a TSV $\delta\mu/e \propto -\frac{dS}{dn} \Delta T$ when a temperature difference $\Delta T$ is built up between spin injector and detector; here it is assumed that $\Delta T$ is independent of $n$.}
\label{Fig4}
\end{figure}

\end{document}